\documentclass[10pt,aps,pra,twocolumn,showpacs,preprintnumbers,amsmath,amsfonts,amssymb,groupedaddress,longbibliography,floatfix]{revtex4-1}

\usepackage[T1]{fontenc}
\usepackage[utf8]{inputenc}

\usepackage[caption=false]{subfig}
\usepackage{graphicx}
\usepackage{dcolumn}
\usepackage{bm}

\usepackage{amsfonts}
\usepackage{amsmath}
\usepackage{amssymb}
\usepackage{mathtools}

\usepackage[]{xcolor}
 \definecolor{lightblue}{RGB}{0,0,255}
 \definecolor{darkblue}{RGB}{0,0,128}
 \definecolor{orange}{RGB}{255,127,0}
 \definecolor{green}{RGB}{128,128,0}
 
\usepackage[
bookmarks=true,
bookmarksopen=false,
bookmarksnumbered=true,
breaklinks=true,
colorlinks=true,
linkcolor=blue,
anchorcolor=black,
citecolor=black,
filecolor=black,
menucolor=black,
urlcolor=black,
pdfpagemode=UseOutlines,
pdftitle=Synthetic dimensions in the strong-coupling limit: supersolids and pair-superfluids,
pdfsubject=Synthetic dimensions in the strong-coupling limit: supersolids and pair-superfluids,
pdfauthor=Thomas Bilitewski
]{hyperref}

\DeclareSymbolFont{bbold}{U}{bbold}{m}{n}
\DeclareSymbolFontAlphabet{\mathbbold}{bbold}

\DeclarePairedDelimiterX\braket[3]{\langle}{\rangle}%
{#1\,\delimsize\vert\,#2\,\delimsize\vert\,#3}

\DeclarePairedDelimiterX\expval[1]{\langle}{\rangle}{#1} 

\DeclareMathOperator{\Tr}{Tr}

\begin{document}

\title{Synthetic dimensions in the strong-coupling limit: supersolids and pair-superfluids}
\author{Thomas \surname{Bilitewski}}
\email{tb494@cam.ac.uk}
\affiliation{T.C.M. Group, Cavendish Laboratory, J.J. Thomson Avenue, Cambridge CB3 0HE, United Kingdom}
\author{Nigel R. \surname{Cooper}}
\affiliation{T.C.M. Group, Cavendish Laboratory, J.J. Thomson Avenue, Cambridge CB3 0HE, United Kingdom}
\date{\today}
\begin{abstract}
We study the many-body phases of bosonic atoms with $N$ internal states confined to a 1D optical lattice under the influence of a synthetic magnetic field and strong repulsive interactions. 
The $N$ internal states of the atoms are coupled via Raman transitions creating the synthetic magnetic field in the space of internal spin states corresponding to recent experimental realisations. 
We focus on the case of strong $\mbox{SU}(N)$ invariant local density-density interactions in which each site of the 1D lattice is at most singly occupied, and strong Raman coupling, in distinction to previous work which has focused on the weak Raman coupling case. This allows us to keep only a single state per site and derive a low energy effective spin $1/2$ model. The effective model contains first-order nearest neighbour tunnelling terms, and second-order nearest neighbour interactions and correlated next-nearest neighbour tunnelling terms. 
By adjusting the flux $\phi$  one can tune the relative importance of first-order and second-order terms in the effective Hamiltonian. In particular, first-order terms can be set to zero, realising a novel model with dominant second-order terms. We show that the resulting competition between density-dependent tunnelling and repulsive density-density interaction leads to an interesting phase diagram including a phase with long-ranged pair-superfluid correlations. 
The method can be straightforwardly extended to higher dimensions and lattices of arbitrary geometry including geometrically frustrated lattices where the interplay of frustration, interactions and kinetic terms is expected to lead to even richer physics. 
\end{abstract}


\maketitle

\section{\label{sec:intro}Introduction}
Cold-atom systems provide an ideal setting in which to perform experiments that simulate quantum many-body systems. The fine control and wide tunability of system parameters, as well as precise measurements of observables, allow cold-atom systems to simulate idealized models of solid state physics\cite{Bloch2008,Ketterle_2008}, making them a testing ground for condensed matter theories. However, the natural situations that emerge in cold atom experiments also introduce new models with interesting and novel features which raise new theoretical questions.

Optical lattices can be used to confine atoms in $d=1,2,3$-dimensional lattices of chosen geometry. An additional synthetic dimension can be created in any $d$-dimensional lattice by exploiting the internal states, e.g. spin states, of the atoms \cite{Boada2012,Celi2014}. Recent progress in the control of cold atomic gases allows the study of systems with large (tunable) numbers of internal spin states \cite{Pagano2014,Krauser2012,Krauser2014}.
Engineering the transitions between the internal states, e.g. by Raman transitions induced by lasers, allows the simulation of motion as along an additional (finite) ``synthetic'' dimension. 
Moreover, this synthetic dimension can also be used to engineer artificial gauge fields for neutral atoms \cite{Dalibard2011,Celi2014,Goldman2014}, thus opening the possibility to explore topological physics in higher dimensional settings.
Experimentally, this has been realized in a 1-dimensional lattice geometry realising the physics and associated topological properties of a 2-dimensional system \cite{Stuhl2015,Mancini2015}.
More recently, it has been proposed how one could simulate 4-dimensional Quantum Hall physics in cold atom setups using these techniques \cite{Price2015}.

In this work, we will focus on one-dimensional systems with a finite synthetic dimension composed of $N=(2 I+1)$ spin states coupled by laser beams in such a way as to create an artificial magnetic field. Thus, they can alternatively be considered as frustrated $N$-leg ladders. Optical lattice experiments with cold atoms motivate the study of both bosonic \cite{Dhar2012,Dhar2013,Tokuno2014,Piraud2015,Greschner2015,Kolley2015} and fermionic systems \cite{Roux2007,Sun2013,Piraud2014,Lacki2015,Budich2015,Mazza2015,Cornfeld2015,Zeng2015,Barbarino2015a,Barbarino2015}. 
The predicted behaviour includes chirally ordered phases \cite{Dhar2013}, vortex phases \cite{Piraud2015}, magnetic crystals and quasi-1D analogues of fractional Quantum Hall states
\cite{Cornfeld2015,Barbarino2015a,Barbarino2015}.
At the centre of these phenomena is the interplay of the gauge fields and the $\mbox{SU}(2I+1)$ symmetric interactions \cite{Manmana2011,Gorshkov2010,Zhang2014,Cazalilla2014}.
The natural $\mbox{SU}(2I+1)$ symmetry of the interactions between the spin states implies, in the interpretation of a ladder, that the interactions are infinitely ranged along the synthetic dimension and short-ranged along the real dimension in contrast to the situation usually considered in the solid-state context.  We remark that therefore the limit of hardcore interactions of bosonic particles does not correspond to a Tonks-Girardeau gas \cite{Girardeau1960,Kinoshita2004,Paredes2004} and the system does not reduce to free fermions.

Prior studies have focused on the weak Raman coupling case in which one obtains helical states and edge currents \cite{Barbarino2015}. In contrast we will study the case of strong Raman coupling and strong interactions, focusing on an effective model of hardcore bosons/spinless fermions in these limits which can alternatively be understood in terms of an effective pseudo spin-$1/2$ system. 
Our main focus will be on a regime in which the physics is dominated by the interplay of density-density interactions and correlated tunnelling terms. This will lead to a competition between phase-separation and charge-order, and normal superfluidity and pair-superfluidity. 

In 2D pair-superfluids can be realised using the long-range interactions of dipolar quantum gases \cite{BARANOV2008,Baranov2012} and confining them in bi-layer geometries \cite{Trefzger2009,Safavi-Naini2013,Macia2014}.
In a mean field analysis the presence of correlated tunnelling allows the condensation of pairs ($\expval{b_i b_j}\ne 0$) in the absence of single-particle condensation ($\expval{b_i}=0$) \cite{Bendjama2005}. Generically, correlated tunnelling can be understood to act as an attractive interaction between the bosons favouring pair formation, and the repulsive nearest neighbour interaction is required to avoid collapse \cite{Nozieres1982} or phase-separation \cite{Schmidt2006}. Correlated tunnelling has been shown to lead to pair-superfluidity for bosons in 2D \cite{Schmidt2006} and in 1D \cite{PSF_Takayoshi_2013,Tovmasyan2013} in theoretical studies, but the required models are hard to realise experimentally.

We propose a way to realise (quasi) pair-condensed and supersolid phases of ultracold atoms starting from an experimentally realised system. We do not require special (long-range) interactions or complicated lattice geometries. The proposed scheme is applicable to both fermions and bosons, but we will limit the discussion to the bosonic case here.
We do not assume specially engineered Raman couplings of the spin states to obtain homogeneous couplings along the synthetic dimension or periodic boundary conditions which are hard to realise experimentally for large number of internal spin states, but consider the highly non-homogeneous couplings and open boundary conditions along the synthetic dimension which occur naturally for $I>1$ due to the nature of the atom-light interaction.

We introduce the full model and the effective model derived in the limits of large Raman coupling and strong interactions in Sec.~\ref{sec:model}.
Importantly, the coupling constants will turn out to depend on the flux $\phi$, and the freedom in tuning both the flux and the number of spin states $2 I+1$ allows great control and  freedom in engineering the resulting effective Hamiltonian.
In Sec.~\ref{sec:model_phi_pi} we will focus on the special case of flux $\phi=\pi$ in which the first order terms vanish and investigate the behaviour resulting from the dominant second order terms in the effective model. By employing Density-Matrix-Renormalization Group (DMRG) calculations \cite{Schollwoeck2005} the phase-diagram of the effective model is obtained, and described in Sec.~\ref{subsec:res_model_phase}.
Based on the analysis of correlation functions and the von-Neumann entropy we establish a phase-diagram containing a charge-density wave (CDW) at half-filling, a supersolid phase (SS) with simultaneaous charge-density wave order and superfluid correlations, and a (quasi) pair-superfluid phase.
\section{\label{sec:model}Model}
We consider spinful bosons with $N=2 I +1$ internal spin states loaded into a one-dimensional optical lattice described by a Hamiltonian
$\hat{H} = \hat{H}_1 +  \hat{H}_{2} + \hat{H}_{\mathrm{ int} }$. 
$\hat{H}_1$ describes the bosonic hopping along the lattice,
$
 \hat{H}_1 = -t \sum_j \sum_{m=-I}^{I} \left( \hat{c}^{\dagger}_{j+1,m} \hat{c}_{j,m} + h.c \right) 
$
where $\hat{c}^{(\dagger)}_{j,m}$ are bosonic operators annihilating (creating) bosons in spin state $m$ at site $j$ and $t$ is the hopping amplitude.
$\hat{H}_2 $ describes the Raman coupling of the internal spin states via
$
 \hat{H}_2 = - \sum_j \sum_{m=-I}^{I-1} \Omega_{m+1} \left( e^{i \phi j} \hat{c}^{\dagger}_{j,m+1} \hat{c}_{j,m} +h.c \right)
$
where $\Omega_m= \Omega g_m $ with $g_m=\sqrt{I(I+1) -m (m-1)} $ and $\phi$ is the running phase of the Raman beams (set by the wavevector transfer $\Delta k$ and the lattice constant $d$).
$\hat{H}_{\mathrm{ int} }$ is taken to be an $\mbox{SU}(2I+1)$ invariant interaction of contact form, i.e. $\hat{H}_{\mathrm{ int} }= U \sum_{j,m,m^{\prime}} \hat{n}_{j,m}( \hat{n}_{j,m^{\prime}}-\delta_{m,m^{\prime}})$.
In the next section we will consider an effective spin-$1/2$ model describing the dynamics in the strong coupling limit.
\subsection{\label{sec:eff_model}Effective Model at strong coupling}
We will consider the parameter regime $ t \ll \Omega, U $ and work with the resulting low-energy effective Hamiltonian in the following. In the limit $t \ll \Omega$ only the lowest of the eigenstates of $\hat{H}_2$ remains in the effective description coupled via direct and virtual hoppings induced by $\hat{H}_1$. The interaction $\hat{H}_{\mathrm{ int} }$ takes the same form in the eigenbasis of $\hat{H}_2$ due to its $\mbox{SU}(2I+1)$-invariance and in the limit of $ t \ll U $ leads to a hardcore constraint in the effective basis.
In App.~\ref{app:derivation} we derive the effective second-order model describing spinless particles interacting via a nearest neighbour interaction and hopping with nearest neighbour, next-nearest neighbour and correlated next-nearest neighbour tunnelling terms.

The effective Hamiltonian takes the form
 \begin{align}
  \hat{H}_{\mathrm{ eff} }/t &=  - t_1(\phi) \sum_j \left(\hat{d}^{\dagger}_{j+1} \hat{d}_j +h.c.\right) +\kappa V(\phi,\tilde{u}) \sum_l \hat{n}_l \hat{n}_{l+1} \notag\\
&\quad -\kappa t_2(\phi,\tilde{u}) \sum_j \left(\hat{d}^{\dagger}_{j+2}\hat{d}_j + h.c\right) \label{eq:H_eff}\\
&\quad +\kappa t_{\mathrm{ cor} }(\phi,\tilde{u})\sum_j \left(\hat{d}^{\dagger}_{j+2} \hat{n}_{l+1} \hat{d}_j + h.c \right) \notag
 \end{align}
where $\hat{d}_{j}=\hat{d}_{j,I}$ is the creation operator for a particle in the $s_x=I$ (after the unitary transformation explained in App.~\ref{app:derivation}) eigenstate at site $j$ , $\kappa =t/\Omega$, and $\tilde{u}=U/(4\Omega I)$.
The explicit form and functional dependence of the coupling constants on the flux $\phi$, the interaction strength $\tilde{u}$ and the number of spin states $I$ is provided in App.~\ref{app:derivation}, Eq.~\ref{eq:app_f_coupling1}-\ref{eq:app_f_coupling3}.

The first term describes the direct hopping between the $s_x=I$ spin state on neighouring lattice sites, with an energy scale that is reduced from the bare hopping $t$ by the factor $t_1(\phi) = (\cos\phi/2)^{2I}$ (Eq.~\ref{eq:app_f_coupling1}).
The remaining terms describe virtual hopping processes, with energy scale proportional to $t \kappa = t^2/\Omega$. The nearest neighbour repulsion $V$ contains three contributions, originating from nearest neighbour hopping and returning to the original site via an excited spin state on a neighbouring site which is either empty or occupied or hopping onto an occupied site in the lowest energy spin state. The correlated tunnelling term $t_{\mathrm{ cor} }$ arises from the corresponding processes with the particle not returning to the original site. These processes are illustrated in Fig.~\ref{fig:effective_model}(a).

Importantly, the virtual hopping between the different Raman eigenstates is controlled by $\kappa =t/\Omega$. To avoid double occupancy we only require $t_1(\phi) \ll U/t$, which can be achieved even if the bare coupling $t$ is large by making $t_1(\phi)$ small through a judicious choice of $\phi$. This allows us to work at relatively high energy scales using shallow lattices with high bare tunnelling rates $t$, in contrast to the induced interactions in the Mott regime of the Hubbard model scaling with $t/U$ requiring deeper lattices and lowering the overall energy scale. 
Further, the dependence of the coupling constants on the flux $\phi$ allows one to eliminate the first order tunnelling terms and obtain an effective model with dominant second order terms even for relatively shallow lattices where all energy scales remain large.
\section{\label{sec:model_phi_pi}Model at $\phi=\pi$} 
\begin{figure} 
\centering 
\includegraphics[width=0.98\linewidth]{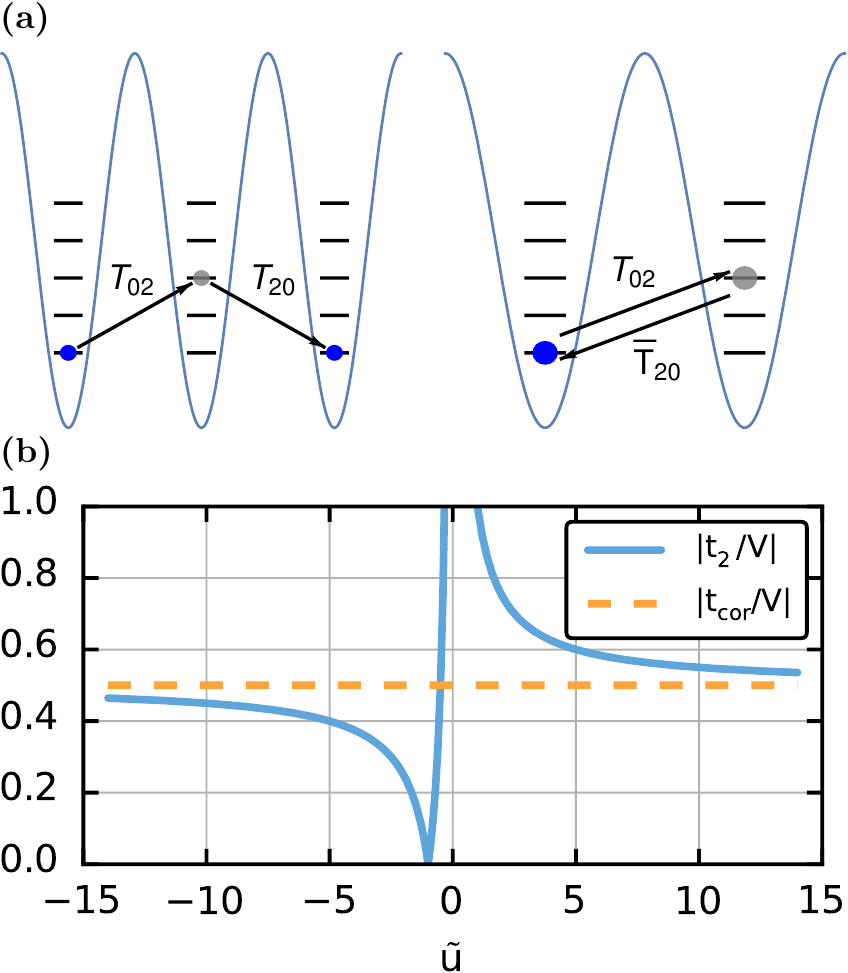}
\caption{(a) Second order virtual proccesses in the effective Hamiltonian Eq.~\ref{eq:H_eff} illustrated in the case of $I=2$. The left shows a particle hopping into an excited state on an unoccupied site and back to the ground state and leads to a normal and a correlated NN neighbour hopping term $t_{2}$ and $t_{\rm cor}$, hopping via an occupied site leads to $t_{\rm cor}$, and the right corresponds to hopping back and forth via an excited state and leads to an effective NN interaction $V$ of particles on neighbouring sites.
(b) Coupling constants $t_2/V$ and $t_{\rm cor}/V$ as a function of rescaled interaction $\tilde{u}=U/(4\Omega I)$ of the effective spin $1/2$ model at flux $\phi=\pi$, see Eq.~\ref{eq:model_phi_pi}
\label{fig:effective_model}}
\end{figure}
In the following we focus on the model at flux $\phi=\pi$. Then, the first order nearest neighbour tunnelling term $t_{1}(\phi)$ vanishes identically and the effective model is determined by the second order terms only. 
The model reduces to
\begin{gather} \label{eq:model_phi_pi} 
\begin{aligned}  
  \hat{H}_{\mathrm{ eff} }/(t \kappa) &=V \sum_{j} \hat{n}_{j} \hat{n}_{j+1}  -t_{2}  \sum_j  \left(  \hat{c}_{j}^{\dagger}   \hat{c}_{j+2}  + h.c.  \right)\\
  &\quad +V/2 \sum_j  \left(  \hat{c}_{j}^{\dagger} \hat{n}_{j+1}   \hat{c}_{j+2 }  + h.c.  \right)
  \end{aligned}
\end{gather}
 where $c_{j}^{\dagger} $ is the creation operator for hard-core bosons or spinless fermions at site $j$ and $n=c_{j}^{\dagger} c_{j}$ the corresponding density and the couplings are the ones defined below Eq.~\ref{eq:H_eff} for $\phi=\pi$. Note that in these limits $t_{\mathrm{cor}}=V/2$.
 Since the NN tunnelling term has dropped out, particles now only hop on their respective $A/B$ sublattices, the model can therefore also be understood to live on a ``zigzag'' lattice.

To gain some understanding of the effective model, we first consider the more general case in which all coupling constants can be tuned independently, i.e. we consider the model with couplings $t_{2}$, $t_{\mathrm{cor}}$ and $V$. Note that those correspond to to 2-body, 3-body and 4-body terms respectively. 
For $t_{\mathrm{cor}}=V=0$ the model is non-interacting and describes free hardcore bosons living separately on each sublattice. 
For $t_{\mathrm{cor}}=0$ the model corresponds to the $t_2-V$-model \cite{Ghosh2014}. It has been shown to undergo a quantum phase-transition from a superfluid (SF) phase to a supersolid (SS) at non-half filling and to a charge-density-wave (CDW) at exactly half-filling as a function of $t_2/V$.
For $V=0$ the model is integrable and known as Bariev's model \cite{Bariev1991}, in this limit we have two NNN hopping terms, a normal hopping $t_2$ and a correlated hopping $t_{\mathrm{cor}}$ for  which hopping between sites depends on the occupation of the intermediate site on the other sublattice. Depending on $t_{\mathrm{cor}}/t_2$ the model has a finite CDW amplitude, i.e. different sublattice populations, in the groundstate.
The fermionic spin $1/2$ version of this model has recently been studied in Ref.~\cite{Chhajlany2016}. For $V=0$ and $t_{\mathrm{cor}}=t_2$, the model admits an exact solution via a mapping to free spinless particles moving on a charge lattice. This solution becomes possible, because for $t_{\mathrm{cor}}=t_2$ particles cannot pass each other, and the sequence of particles remains preserved throughout the dynamics. The groundstate of the model is found to be a paired-hole superconductor with hidden string order and algebraically decaying 2-particle correlations.
 
For our model, we are not free to choose these couplings independently. The dependence of the couplings in the effective model Eq.~\ref{eq:model_phi_pi} on the rescaled interaction strength $\tilde{u}=U/(4\Omega I)$ is shown in Fig.~\ref{fig:effective_model}(b). In these limits we obtain $t_{\mathrm{ cor} }/V=0.5$ and $t_2/V=(1+\tilde{u})/(2\tilde{u})$.
Thus, the model depends only on a single free parameter, $\tilde{u}$, which determines the ratio $t_{2}/V$, or we can alternatively consider the model as a function of $t_{2}/V$. Hardcore interactions correspond to $t_{2}/V =0.5$ and we will consider the region of repulsive interactions corresponding to $t_{2}/V \ge 0.5$ in Sec.~\ref{subsec:res_model_phase}. We note that with these parameters we are outside of the integrable limits described above and it will be interesting to see what remains of the physics in the parameter regime accessible in our model.
\subsection{Phase Diagram\label{subsec:res_model_phase}}
\begin{figure} 
\centering 
\includegraphics[width=0.98\linewidth]{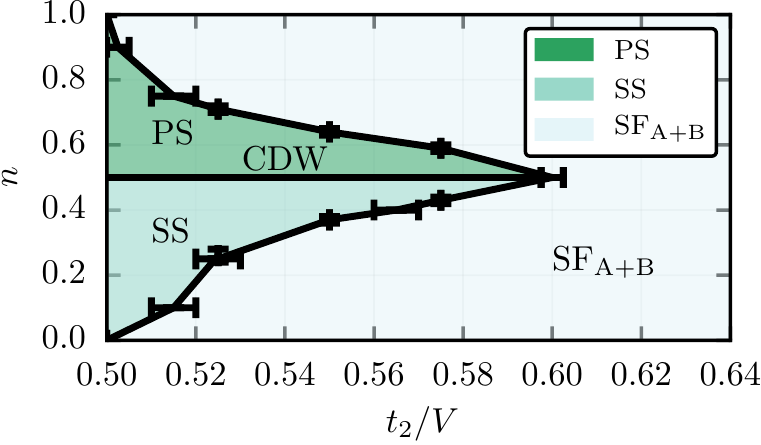}
\caption{Phase-diagram of the effective model Eq.~\ref{eq:model_phi_pi} obtained from the DMRG calculations as a function of coupling $t_2$ and density $n$. Three distinct phases are observed, a gapped CDW at $n=0.5$ and $t_{2} \le 0.6$, a supersolid phase ($\mbox{SS}$) with superfluid order on one of the sublattices with the other sublattice being empty with central charge $c=1$ below half-filling $n < 0.5$, and a homogeneous phase with dominant superfluid ($\mbox{SF}_{\mathrm{ A+B}}$) order on both lattices with $c=2$ for high densities $n$ and high $t_{2}$ which also shows strong pair-superfluid correlations. Above half-filling at low $t_{2}$ we find phase-separation (PS) as indicated in a jump in $n(\mu)$.
\label{fig:DMRG_phase}}
\end{figure}
To characterise the ground state phases we perform DMRG simulations using the ALPS MPS framework \cite{ALPS,Dolfi2014}.
We consider system sizes of $L= 80,120,160,240$ with open boundary conditions keeping a maximal number of states of $m=400,600,800$, extrapolating results for fixed system size in $1/m$.
To characterize the ground state and obtain the phase-diagram we study two- and four-point correlation functions and the structure factors for CDW, superfluid and pair-superfluid order. To reduce the effects of the open boundary conditions correlators are measured from the middle of the system and averaged over a window of 10 sites around the central site. We perform finite-size scaling of the corresponding correlation-lengths, decay exponents and structure factors to obtain the phase-boundaries.
In addition we characterise the phases via their entanglement entropy and central charge.

On a bipartite lattice, due to the vanishing of the nearest neighbour tunnelling, the sublattice populations $n_{A(B)}=\sum_{i \in A(B)} n_{i}$ are separately conserved, and we focus on equal populations on both sublattices $n_A=n_B$.
The phase diagram of the model as a function of $t_{2}$ in the range $0.5 \le t_{2}/V \le 0.64$ and density $0 \le n \le 1$ is shown in Fig.~\ref{fig:DMRG_phase}. Three distinct phases are observed in this parameter range, a charge-density wave (CDW) with a period of two lattice sites, a supersolid ($\mbox{SS}$) with simultaneous (quasi-) superfluid and maximal CDW order, and a homogeneous phase with (quasi-) superfluidity on both sublattices ($\mbox{SF}_{\mathrm{ A+B}}$) with pair-superfluid correlations.
Since we consider the case of $n_A=n_B$, both the CDW and the $\mbox{SS}$ phases are additionally separated into a left/right region with vanishing density on one of the sublattices in both regions. 
Before the transition into the $\mbox{SF}_{\mathrm{ A+B}}$-phase, the maximally imbalanced state with $N_A=N$, $N_B=0$ (or the equivalent state with $N_A=0$, $N_B=N$) is slightly lower in energy, and degenerate in the thermodynamic limit.
Both the balanced and the maximally imbalanced state are realised as thermodynamic phases when introducing two chemical potentials $\mu_A$ and $\mu_B$ coupling to the respective densities. Our discussion of the properties of the state does not rely on this distinction.
After the transition into the $\mbox{SF}_{\mathrm{ A+B}}$ the imbalanced state is energetically disfavoured.  

Exactly at half-filling $n=0.5$ the CDW phase is stabilised and persists up to $t_2/V=0.6$. Below half-filling $n<0.5$ at low coupling $t_2$ the effects of the nearest neighbour repulsion are still dominant, resulting in a phase where one of the sublattices is empty and the other is filled and becomes (quasi-)superfluid, thus forming a supersolid state. We remark that if one sublattice is empty, the model reduces to free particles hopping on the other sublattice with amplitude $t_2$. 
Above half-filling $n >0.5$ at low $t_2$ particles cannot avoid the cost of the interaction energy $V$ and the system phase-separates. At sufficiently high $t_2$ the effect of the repulsion $V$ can be overcome and a homogeneous phase with superfluid order on both lattices emerges. In this regime all of $t_2$, $t_{\mathrm{ cor} }$ and $V$ are relevant.
\begin{figure} 
  \centering 
  \includegraphics[width=0.98\linewidth]{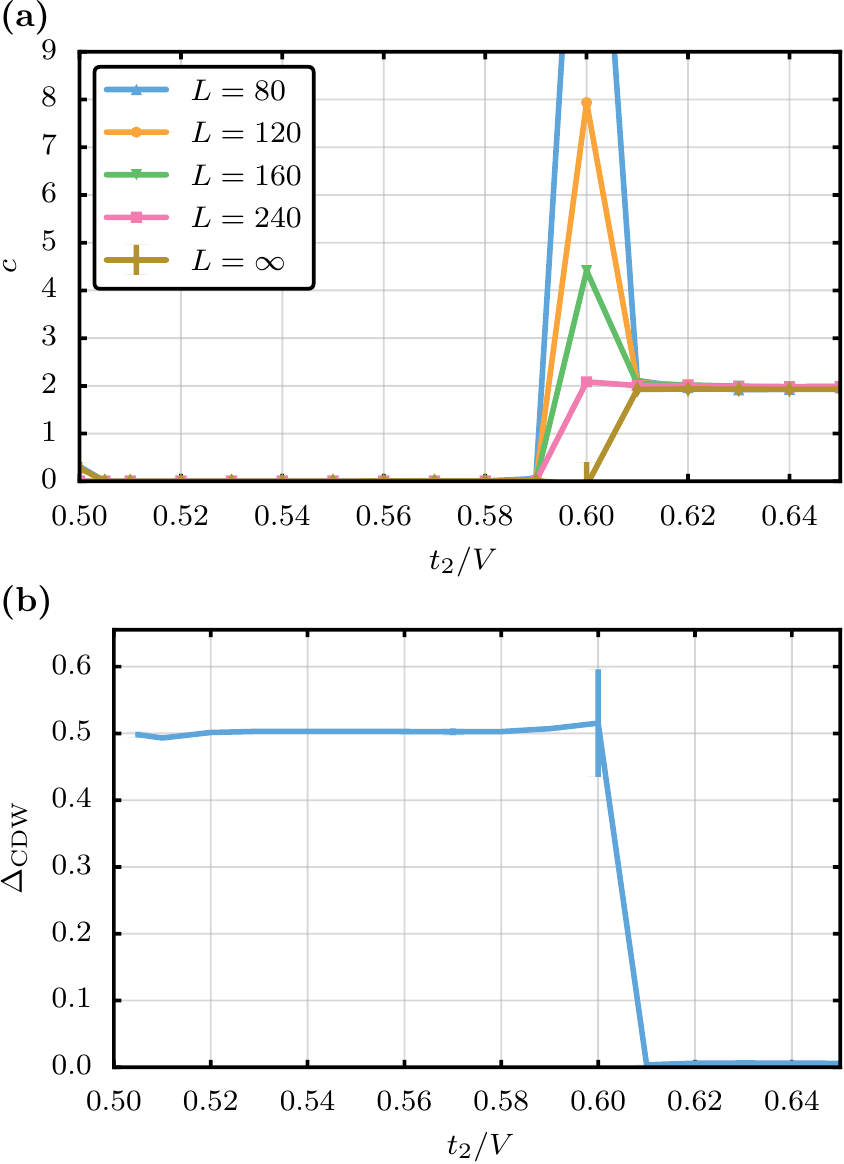}
  \caption{(a) The central charge $c$ determined from fitting the von Neumann block entropy via $S^N_L(l)= \frac{c}{6} \ln \left[\frac{2 L}{\pi} \sin \left( \frac{\pi l}{L}\right) \right]$ for different system sizes as a function of the coupling $t_2$ at density $n=0.5$. The CDW is gapped and the transition occurs into the $\mbox{SF}_{\rm A+B}$ phase with central charge $c=2$.
  (b) Extrapolated CDW order parameter $\Delta_{\rm CDW} = \lim_{L \rightarrow \infty}\sqrt{|1/L\sum_l e^{i\pi l} G(l)|}$ at density $n=0.5$ as a function of coupling $t_2$ showing the vanishing of CDW order at $t_2/V=0.61$.
  \label{fig:central_charge_CDW}}
\end{figure}

An important tool to characterise the ground state behaviour of strongly correlated systems in one dimension is the von Neumann block entropy \cite{Entropy_review}. This is defined as $S^N_A=\Tr \rho_A \ln \rho_A$ where $\rho_A$ is the reduced density matrix $\rho_A ={\Tr}_B \rho$ obtained by dividing the chain into the block $A$ consisting of sites $i=1, \cdots, l$ and $B$ of sites $i=l+1,\cdots,L$.
In particular, for a gapped state the entropy saturates whereas it diverges for a gapless state \cite{Kitaev_2003,Korepin_2004}.
For a 1D system of size $L$ with open boundary conditions the von Neumann block entropy behaves as $S^N_L(l)=s_1+\frac{c}{6} \ln \left[\frac{2 L}{\pi} \sin \left( \frac{\pi l}{L}\right) \right]$ where $c$ is the central charge of the associated CFT and $s_1$ is a non-universal constant \cite{Affleck_1991,Holzhey_1994,Calabrese_Cardy_2004}. By fitting $S^N_L$ linearly in the conformal distance $\lambda=\ln \left[\frac{2 L}{\pi} \sin \left( \frac{\pi l}{L}\right)\right]$ we obtain the central charge $c$ of the phase.
The behaviour of the central charge $c$ as a function of the coupling $t_2$ at density $n=0.5$ is shown Fig.~\ref{fig:central_charge_CDW}(a).
The results indicate a transition close to $t_2/V=0.6$. The state for $t_2/V \le 0.6$ is gapped as expected for the CDW phase and the transition occurs into a state with with central charge of $c=2$ in the $\mbox{SF}_{\mathrm{ A+B}}$. Finally below half-filling we find a central charge $c=1$ (not shown), which is consistent with superfluidity on one of the sublattices in the $\mbox{SS}$ phase.

The CDW order can be directly extracted from the density-density correlation and the static structure factor. We measure
$G(l) = \expval{\hat{n}_i \hat{n}_{i+l}}$.
The static structure factor is defined as $S_L(q)=1/L\sum_l e^{iql} G(l)$. The CDW order parameter is given by the square root of the structure factor at $q=\pi$, $\Delta_{\mathrm{CDW}}(L)=\sqrt{|S_L(\pi)|}$ and its infinite system size limit, $\Delta_{\mathrm{CDW}} = \lim_{L \rightarrow \infty}\sqrt{|S_L(\pi)|}$.
The finite system results $\Delta_{\mathrm{CDW}}(L)$ are extrapolated via a quadratic fit in $1/L$ to infinite system size.
The results of this extrapolation are shown in Fig.~\ref{fig:central_charge_CDW}(b).
The CDW order parameter vanishes at $t_2/V=0.61$ signalling the transition into the superfluid state.

To characterize the degree of (quasi-)superfluid order we consider the following two point correlation function
 $C_{\alpha}(2l) =\expval{\hat{c}^{\dagger}_{2 i+\alpha} \hat{c}^{\phantom{\dagger}}_{2i+\alpha+2l}}$
on either sublattice ($\alpha =0,1)$.
 \begin{figure} 
 \centering 
 \includegraphics[width=0.98\linewidth]{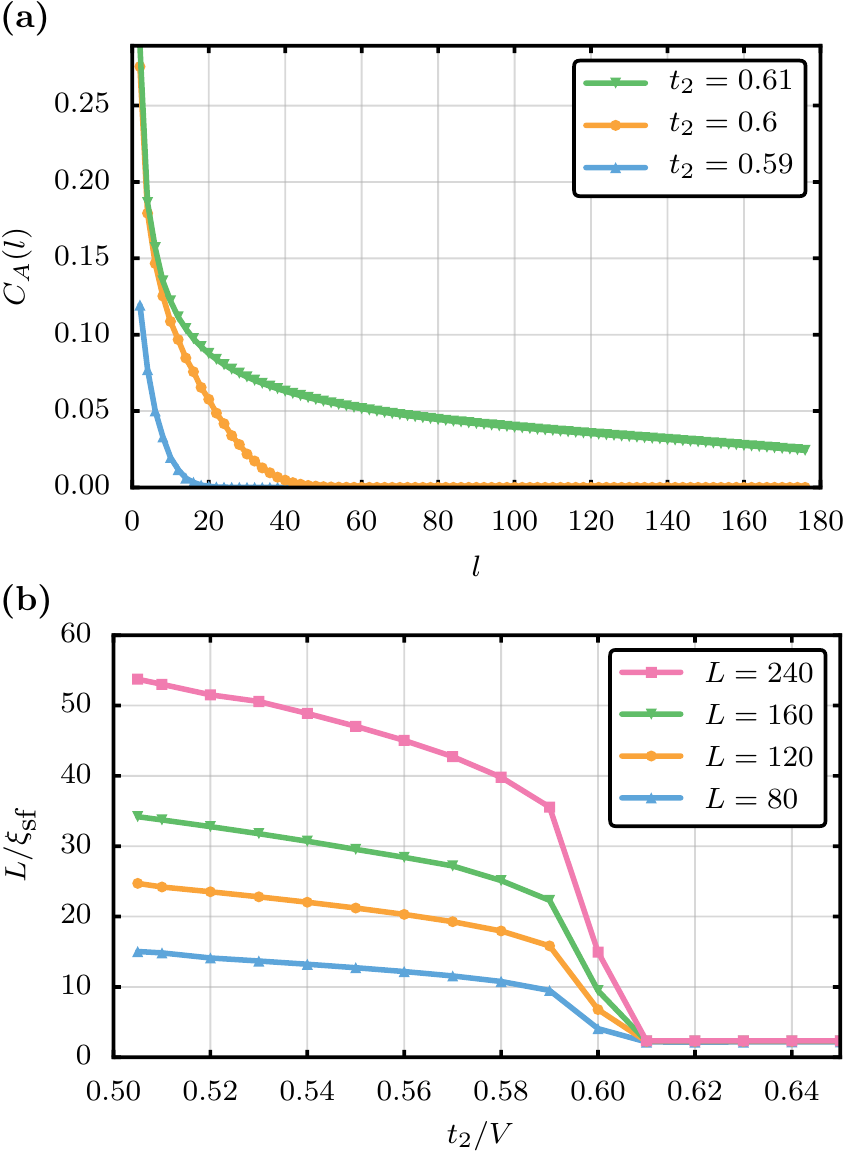}
 \caption{(a) Two-point correlation function $C_{\alpha}(l)$ as a function of $l$ on sublattice A ($\alpha=0$) for a system of size $L=240$ at density $n=0.5$ for $t_2/V=0.61,0.6,0.59$ (top to bottom) showing the transition from short-ranged to long-ranged correlations at $t_2/V=0.61$.
 (b) System size $L$ divided by superfluid correlation length $\xi_{\textrm{sf}}$ for sublattice A vs coupling $t_2$ for $L=240,160,120,80$ (top to bottom). Coalescence of data points for different $L$ at $t_2/V=0.61 \pm 0.05$ signals transition to SF state.
 \label{fig:SF}}
 \end{figure}
This correlation function is shown in Fig.~\ref{fig:SF}(a) for a system of size $L=240$ on sublattice A ($\alpha=0$) displaying a transition from short-ranged to long-ranged correlations around $t_2/V = 0.61$; the other sublattice (not shown) exhibits the same behaviour.
In contrast to CDW there is no order parameter for superfluidity in one dimension, and the whole superfluid phase is critical. Still, the superfluid phase is characterised by a diverging correlation length \cite{White_2000}.  
To determine the transition point we perform finite-size scaling of the correlation length defined as $\xi_{\textrm{sf}}=\sqrt{\sum_l l^2 C_{\alpha}(l)/\sum_l C_{\alpha}(l)}$ \cite{Das_2008,Pandit_2005}. 
In Fig.~\ref{fig:SF}(b) $L/\xi_{\textrm{sf}}$ on sublattice A $(\alpha=0)$ is shown as a function of $t_2$ at $n=0.5$ for different system sizes, the correlations on sublattice B show the same behaviour. The coalescence of the data signals the transition to the superfluid state at $t_2/V=0.61$. 
 \begin{figure} 
 \centering 
 \includegraphics[width=0.98\linewidth]{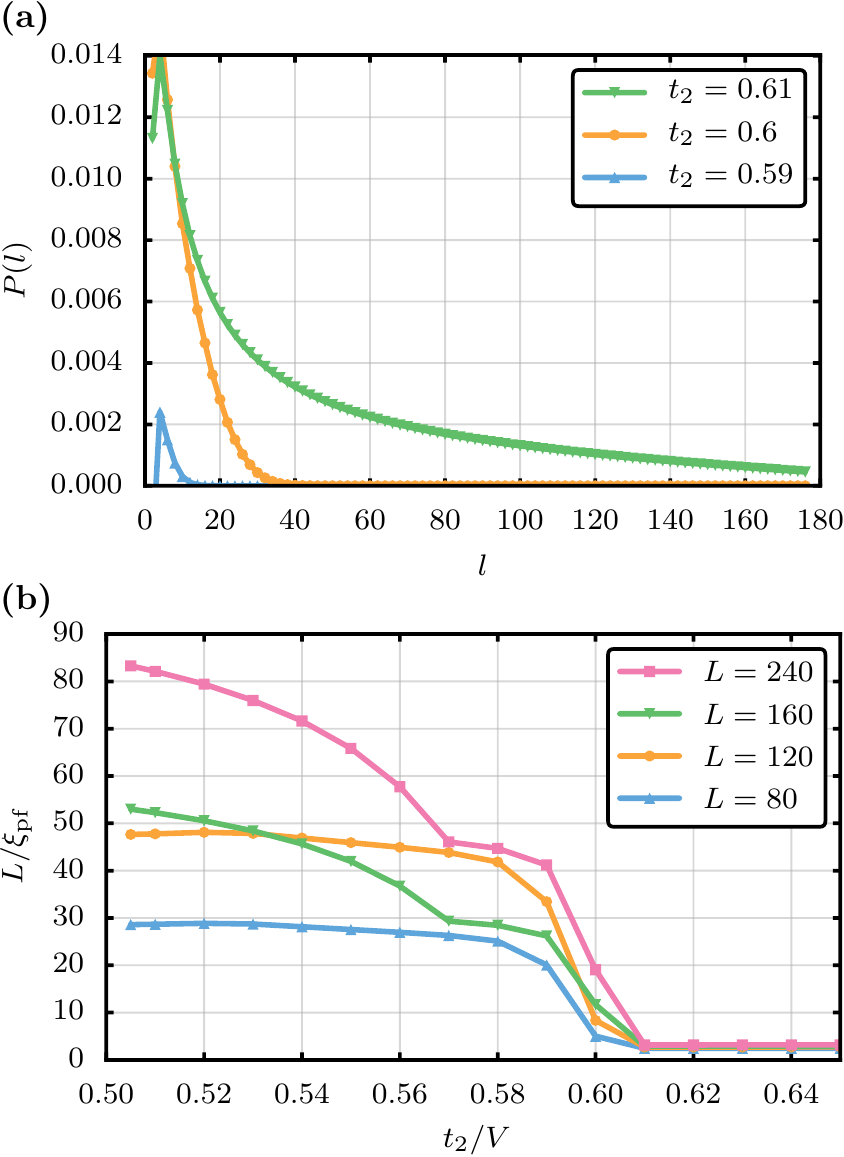}
 \caption{(a) Four-point correlation function $P(l)$ as a function of $l$ for a system of size $L=240$ at density $n=0.5$ for $t_2/V=0.61,0.6,0.59$ (top to bottom) showing the transition from short-ranged to long-ranged correlations at $t_2/V=0.61$.
 (b) System size $L$ divided by pair-superfluid correlation length $\xi_{\textrm{pf}}$ vs coupling $t_2$ for $L=240,160,120,80$ (top to bottom). Coalescence of data points for different $L$ at $t_2/V=0.61 \pm 0.05$ signals transition to PSF state.
 \label{fig:PF}}
 \end{figure}
In the superfluid phase we find strong correlations between the superfluids on the sublattices.
To characterise this phase further we also consider possible condensation of pairs via
$P(2l) =\expval{ \hat{c}^{\dagger}_{2i} \hat{c}^{\dagger}_{2i+1} \hat{c}^{\phantom{\dagger}}_{2i+2l} \hat{c}^{\phantom{\dagger}}_{2i+1+2l}} -\expval{\hat{c}^{\dagger}_{2i} \hat{c}^{\phantom{\dagger}}_{2i+2l}}\expval{\hat{c}^{\dagger}_{2i+1} \hat{c}^{\phantom{\dagger}}_{2i+1+2l}}$ and its corresponding correlation length $\xi_{\textrm{pf}}=\sqrt{\sum_l l^2 P(l)/\sum_l P(l)}$.
The pair-superfluid correlator is shown in Fig.~\ref{fig:PF}(a) and the finite size scaling of the correlation length in Fig.~\ref{fig:PF}(b). We observe very strong pair-superfluid correlations in the $\mbox{SF}_{\mathrm{ A+B}}$ phase consistent with quasi-condensation of pairs as the system becomes superfluid. However, single-particle superfluidity persists alongside pair-superfluidity in the parameter regime we have studied.

\begin{figure*} 
 \centering 
 \includegraphics[width=0.98\linewidth]{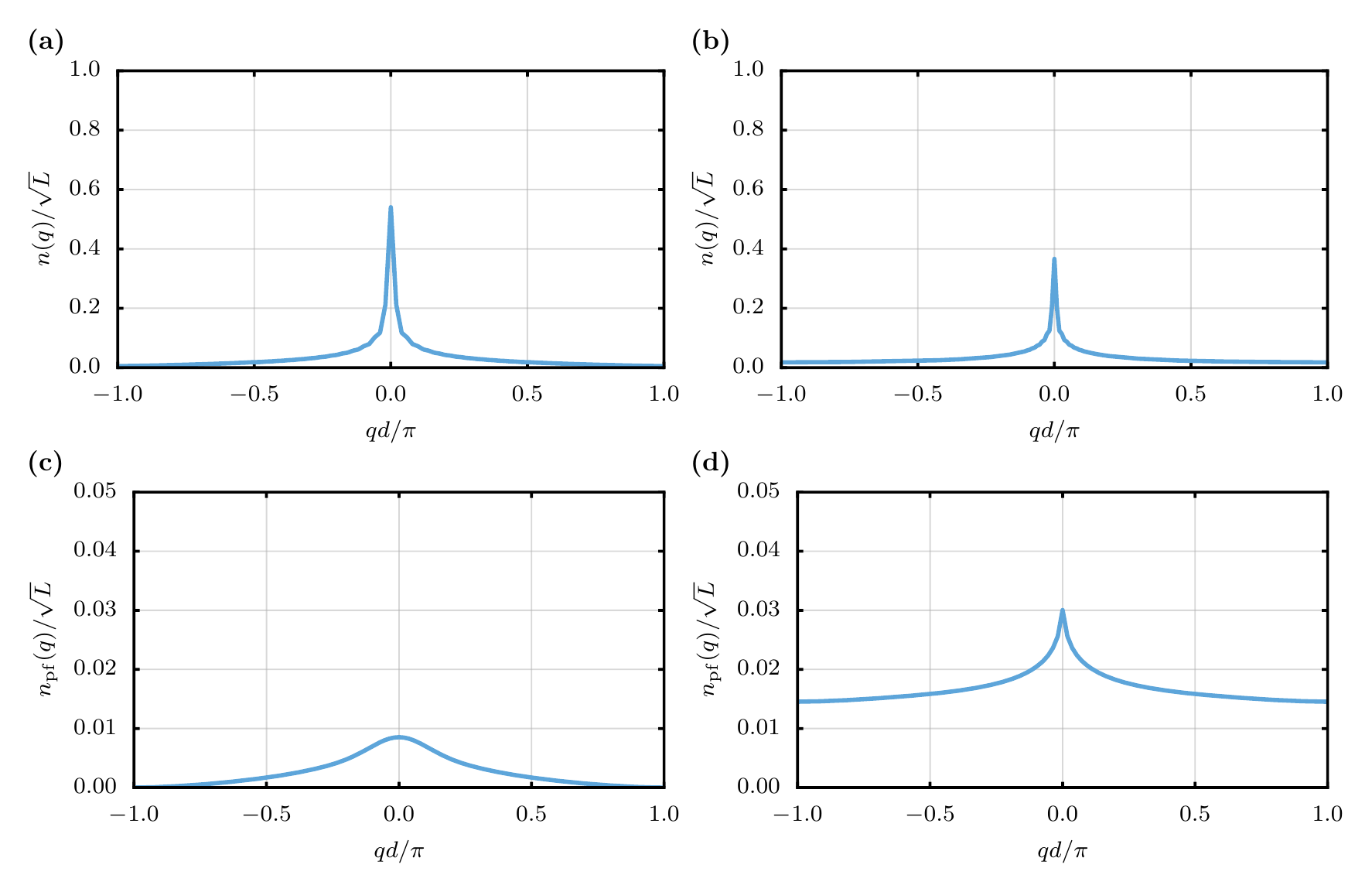}
 \caption{(a)-(b) Single-particle momentum-distribution $n(q)=\sum e^{i q l} C_A(l)$ on sublattice A. (c)-(d) Momentum distribution for pairs of particles $n_{\rm pf}(q)=\sum e^{i q l} P(l)$.
 All at density $n=0.25$ and (a) and (c) at $t_2/V=0.51$ in the $\mbox{SS}$ phase and (b) and (d) at $t_2/V=0.55$ in the $\mbox{SF}_{\mathrm{ A+B}}$ phase. The single-particle momentum distribution shows a quasi-coherent peak in both phases (a) and (b). In contrast, for pairs in the $\mbox{SS}$ phase in (c) no quasi-coherent peak is observed, whereas a peak forms in the $\mbox{SF}_{\mathrm{ A+B}}$ phase in (d).
 \label{fig:momentum_dist}}
\end{figure*}
In Fig.~\ref{fig:momentum_dist}(a) and (b) we display the momentum distribution of particles on sublattice $A$, $n(q)=\sum e^{i q l} C_A(l)$, and in (c) and (d) the momentum distribution of pairs of particles $n_{\rm pf}(q)=\sum e^{i q l} P(l)$.
As for hardcore bosons $n(q=0)$ is expected to scale with $\sqrt{L}$ \cite{Rigol2005}, both quantities are normalised by this factor.
We focus on the transition from the $\mbox{SS}$ phase in (a) and (c) to the $\mbox{SF}_{\mathrm{ A+B}}$ phase in (b) and (d). Whereas in the single-particle momentum distribution a quasi-coherent peak is observed in both phases, pairs only quasi-condense in the $\mbox{SF}_{\mathrm{ A+B}}$ phase as seen in (d).

\section{\label{sec:conclusions}Conclusions}
In summary, in this work we have shown that the interplay of (synthetic) gauge fields and interactions in ultracold gas systems leads naturally to effective Hamiltonians with correlated hopping terms.
We start from an experimentally feasible set-up for the creation of artificial magnetic field using synthetic dimensions. We consider this model in the limits of strong Raman coupling of the spin states and strong interactions where it reduces to an effective model with first order nearest neighbour tunnelling, and second order next-nearest neighbour correlated tunnelling terms and nearest neighbour repulsion.
Importantly, the additional degree of freedom given by adjusting the flux $\phi$ allows to engineer effective models dominated by second-order processes with large energy scales.

By working at flux $\phi=\pi$, the first order nearest neighbour tunnelling term is eliminated, and we obtain a novel model with dominant second-order terms. This is a natural route to a large density-dependent tunnelling term, so the proposed scheme is directly relevant to the realisation and study of models with interaction-assisted hopping and kinetic frustration \cite{Eckholt2009,Huber2010,Keilmann2011,Moeller2012,Liberto2014,Kourtis2015,Mishra2016}.

The physics of our effective model involves the competition between the correlated tunnelling which favours pair formation, and the nearest neighbour repulsion which favours local CDW order. We find three distinct phases:  a CDW phase; a supersolid ($\mbox{SS}$) with simultaneous quasi-superfluidity on either sublattice and maximal CDW order; and a quasi-superfluid on both sublattices with strong pair-superfluid correlations $\mbox{SF}_{\mathrm{ A+B}}$.

The model can be directly generalised to fermionic species and higher dimensional lattices of arbitrary geometry. In the case of fermions, the study of attractive interactions seems particularly relevant for the study of paired phases. The extension to higher dimensions promises even more interesting physics, e.g. BKT transitions to novel superconducting states and geometrically frustrated magnetism. We reserve the discussion of the resulting phases for future work.

\acknowledgments
The DMRG calculations have been performed using the ALPS libraries \cite{ALPS,Dolfi2014}. T.B. would like to thank A. L{\"a}uchli for helpful discussions and comments. T.B. acknowledges insightful correspondence with H. Katsura who pointed out the relation of our effective model to the integrable Bariev-model in 1D. This work was supported by EPSRC Grant No. EP/K030094/1.
Statement of compliance with EPSRC policy framework on research data: All data accompanying this publication are directly available within the publication.

\bibliography{all_cap.bib}{}

\appendix
\section{Derivation of effective Model\label{app:derivation}}
We start from the Hamiltonian of bosons with $N=2 I +1$ internal spin states loaded into a one-dimensional optical lattice described by 
$\hat{H} = \hat{H}_1 +  \hat{H}_2 + \hat{H}_{\mathrm{ int}} $. 

$\hat{H}_1$ describes the bosonic hopping along the lattice,
\begin{equation}
 \hat{H}_1 = -t \sum_j \sum_{m=-I}^{I} \left( \hat{c}^{\dagger}_{j+1,m} \hat{c}_{j,m} + h.c. \right) 
\end{equation}
where $\hat{c}^{(\dagger)}_{j,m}$ are bosonic operators annihilating (creating) bosons in spin state $m$ at site $j$ and $t$ is the hopping amplitude.

In addition the internal spin states are coupled by Raman lasers described by the Hamiltonian
\begin{equation}
 \hat{H}_2 = - \sum_j \sum_{m=-I}^{I-1} \Omega_{m+1} \left( e^{i \phi j} \hat{c}^{\dagger}_{j,m+1} \hat{c}_{j,m} +h.c. \right)
\end{equation}
where $\Omega_m= \Omega g_m $ with $g_m=\sqrt{I(I+1) -m (m-1)} $ and $\phi = \Delta k_{R} d$ is the running phase of the Raman beams given by the wave-vector transfer $\Delta k_{R}$ and the lattice spacing $d$.
$\hat{H}_{\mathrm{ int}}$ is taken to be a $\mbox{SU}(2I+1)$ invariant interaction of contact form, i.e. $\hat{H}_{\mathrm{ int}}= U \sum_{j,m,m^{\prime}} \hat{n}_{j,m}( \hat{n}_{j,m^{\prime}}-\delta_{m,m^{\prime}})$.

For open boundary conditions in the synthetic direction using the unitary transformation $\hat{U}$ defined by $\hat{U} \hat{c}_{j,m} \hat{U}^{\dagger}  = e^{i \phi m j} \hat{c}_{j,m}$ the Hamiltonian is transformed to
\begin{align}\label{eq:app_unit_trans}
 \begin{aligned}
  \hat{U}  \hat{H} \hat{U}^{\dagger}   &= -t \sum_j \sum_{m=-I}^{I-1} \left( e^{-i \phi m} \hat{c}^{\dagger}_{j+1,m} \hat{c}_{j,m} + h.c. \right) \\
   & \quad -  \sum_j \sum_{m=-I}^{I-1} \Omega_m\left(  \hat{c}^{\dagger}_{j,m+1} \hat{c}_{j,m} +h.c. \right) +\hat{H}_{\mathrm{ int}}
 \end{aligned}
\end{align}
As we consider $t \ll \Omega $ we now transform to the eigenstates of the Raman coupling Hamiltonian $\hat{H}_2$. After the unitary transformation this is just
$\hat{H}_2 = -2 \Omega \sum_j \hat{\mathcal{S}}_{x,j}$, where $\hat{\mathcal{S}}_{x,j}$ is the  $\hat{\mathcal{S}}_x$ operator for spin $I$ for particles at site $j$. Note in particular that it is now site-independent due to gauging the Raman phase into the hopping part of the Hamiltonian. Consequently, the eigenfunctions are just the $s_x$ eigenstates and the spectrum at each site is given by $E_s = -2 \Omega s$ with $s=-I,\dots,I$.  
Due to the gauge-transformation we performed, this actually corresponds to a rotating spin-orientation in the original basis.

$\hat{H}_1$ in the new basis reads as $\hat{H}_1 = -t \sum_{s,s^{\prime}} \left( T_{s,s^{\prime}}(\phi) \hat{d}^{\dagger}_{j+1,s^{\prime}} \hat{d}_{j,s} +h.c.\right)$
where $\hat{d}^{\dagger}_{j+1,s^{\prime}}$ creates a particle in the $s^{\prime}_x$ eigenstate at site $j$ and we defined the hopping matrix $T_{s,s^{\prime}}(\phi) = \braket{s_x}{e^{-i \phi \hat{\mathcal{S}}_z}}{s^{\prime}_x}$ which now couples states $s$ and $s^{\prime}$.
As the interaction Hamiltonian is $\mbox{SU}(2I+1)$ invariant it takes the same form in the transformed basis, $\hat{H}_{\mathrm{ int}}= U \sum_{j,s,s^{\prime}} \hat{n}_{j,s}( \hat{n}_{j,s^{\prime}}-\delta_{s,s^{\prime}})$ where now the sum runs over the $s_x$ eigenstates. In the limit of strong interactions this restricts the occupation at each site to be $0$ or $1$.

We see that $\hat{H}_2+\hat{H}_{\mathrm{ int}}$ is diagonal in the occupation number basis of $s_x$ eigenstates. In the limit $t \ll \Omega, U $  we treat $\hat{H}_1$ as a perturbation and derive an effective model keeping only the lowest energy eigenstate at each site, i.e. the $s=I$ state, and consider the sector with empty and singly occupied sites. To second order we obtain a model describing spinless particles interacting via a nearest neighbour interaction and hopping with nearest neighbour, next-nearest neighbour and correlated next-nearest neighbour tunnelling terms. The effective Hamiltonian takes the form
\begin{align}
 \hat{H}_{\mathrm {eff}}/t &= - f_t^{I}(\phi) \sum_j \left(\hat{d}^{\dagger}_{j+1} \hat{d}_j +h.c.\right) \notag  \\
                        &+ 2\kappa \left[ f_{V}^{I}(\phi,\tilde{u}=0) -f_{V}^{I}(\phi,\tilde{u})- \frac{f_t^{I}(\phi)^2}{2I\tilde{u}} \right] \sum_l \hat{n}_l \hat{n}_{l+1}  \notag  \\
		       & +\kappa \Big[ f_{\mathrm{cor}}^{I}(\phi,\tilde{u}=0) \sum_j \left(\hat{d}^{\dagger}_{j+2} (1-\hat{n}_{l+1}) \hat{d}_j + h.c. \right) \\
		       &+ f_{\mathrm{cor}}^{I}(\phi,\tilde{u}) \sum_j \left(\hat{d}^{\dagger}_{j+2} \hat{n}_{l+1} \hat{d}_j + h.c. \right)  \notag \\
		       &  \quad -  \frac{f_t^{I}(\phi)^2}{2I\tilde{u}}   \sum_j\left( \hat{d}^{\dagger}_{j+2} \hat{n}_{l+1} \hat{d}_j  + h.c.\right)\Big] \notag 
\end{align}
where $\hat{d}_{j}=\hat{d}_{j,I}$ is the creation operator for a particle in the $s_x=I$ eigenstate at site $j$, $\kappa =t/\Omega$ and $\tilde{u}=U/(4I\Omega)$.

The functions  $f^{(I)}_i(\phi)$  depend on the flux $\phi$, the interaction strength $\tilde{u}$ and parametrically on the number of spin states $I$. The first term describes the diagonal hopping between the $s=I$ spin states and the remaining terms describe virtual hopping processes.
The nearest neighbour repulsion $V$ originates from nearest neighbour hopping and returning to the original site via an excited spin state on a neighbouring site which is either empty (first term) or occupied (second term) or hopping onto an occupied site in the lowest energy spin state (third term). The correlated tunnelling term $t_{\mathrm {cor}}$ arises from the corresponding processes with the particle not returning to the original site.
The functions $f^{(I)}_i(\phi)$ take the explicit form
\begin{align} 
 f_t^{I}(\phi) &= T_{II} (\phi) = \cos(\phi/2)^{2I} \label{eq:app_f_coupling1}\\
 f_{\mathrm{cor}}^{I}(\phi,\tilde{u}) &= -\sum_{s^{\prime} \ne I} \frac{T_{I,s^{\prime}}(\phi) T_{I,s^{\prime}}(\phi)}{(E_{s^{\prime}}-E_I+U)/\Omega} \label{eq:app_f_coupling2}\\
  &=-\frac{\cos(\phi/2)^{4I}}{4 I\tilde{u}} \left[F(-2I,2I \tilde{u},1+2I\tilde{u},\tan(\phi/2)^2) -1\right] \notag \\
 f_{V}^{I}(\phi,\tilde{u}) &= \sum_{s^{\prime} \ne I} \frac{T_{I,s^{\prime}}(\phi) \bar{T}_{s^{\prime},I}(\phi)}{(E_{s^{\prime}}-E_I +U)/\Omega} \label{eq:app_f_coupling3} \\
 &=\frac{\cos(\phi/2)^{4I}}{4 I\tilde{u}} \left[F(-2I,2I\tilde{u},1+2I\tilde{u},-\tan(\phi/2)^2) -1\right] \notag
\end{align}
where $\tilde{u}=U/(4 I\Omega)$ and $F(a,b,c,z)={\prescript{}{2}{F}_1}(a,b,c,z)$ is the hypergeometric function
\end{document}